\documentclass{moriond}

\usepackage{lineno}

\def\Journal#1#2#3#4{{#1} {\bf #2}, #3 (#4)}

\def\PLB{{\em Phys. Lett.}  B}
\def\PRD{{\em Phys. Rev.} D}

\newcommand{\Photo}{}

\begin{document}
\vspace*{4cm}
\title{Observation of $\tau$ lepton pair production in ultraperipheral nucleus-nucleus collisions with the CMS experiment and the first limits on $(g-2)_\tau$ at the LHC}

\author{Arash Jofrehei on behalf of the CMS Collaboration}

\address{Physics Institute, University of Zurich,\\
Zurich, Switzerland}

\maketitle\abstracts{
The first observation of $\tau$ lepton pair production in ultraperipheral nucleus-nucleus collisions, a pure quantum electrodynamics (QED) process, is presented. The measurement is based on a data sample collected by the CMS experiment at a per nucleon center-of-mass energy of $5.02~\mathrm{TeV}$, and corresponding to an integrated luminosity of $404~\mu\mathrm{b}^{-1}$. The $\gamma\gamma\to\tau^{+}\tau^{-}$ production is observed with a statistical significance of at least five standard deviations for $\tau^{+}\tau^{-}$ events with a muon and three charged hadrons in the final state. The cross section is measured in a fiducial phase space region, and is found to be $\sigma(\gamma\gamma\to\tau^{+}\tau^{-}) = 4.8\pm 0.6\,(\mathrm{stat})\pm 0.5\,(\mathrm{syst})~\mu\mathrm{b}$, in agreement with leading-order QED predictions. The measurement, based on a small fraction of the expected integrated luminosity of the LHC nuclear program, establishes the potential for a substantially more precise determination of the anomalous magnetic moment of the $\tau$ lepton, which is currently poorly constrained.}

Ultraperipheral collisions (UPC) of nuclei where the impact parameter is greater than the sum of their radius provide an extremely clean environment to study various photon-induced processes~\cite{Baltz:2008}.
Lead-lead (PbPb) UPC receive an enhancement of $Z^4$ (where $Z=82$) in the cross section of two-photon production processes relative to proton-proton collisions. Recent theoretical studies~\cite{Beresford:2019gww,Dyndal:2020yen} proposed that the production cross section and kinematics of $\tau$ lepton pairs produced in PbPb UPC can be exploited in a novel way. Specifically, the electromagnetic couplings of the $\tau$ lepton, e.g., its anomalous magnetic moment $a_\tau = (g-2)_{\tau} / 2$, can be probed for the first time at LHC, hence allowing fundamental tests of quantum electrodynamics (QED) and probing beyond the standard model physics~\cite{Crivellin:2021spu}.

\begin{figure}[htp]
    \centering
    \includegraphics[keepaspectratio,width=0.55\textwidth]{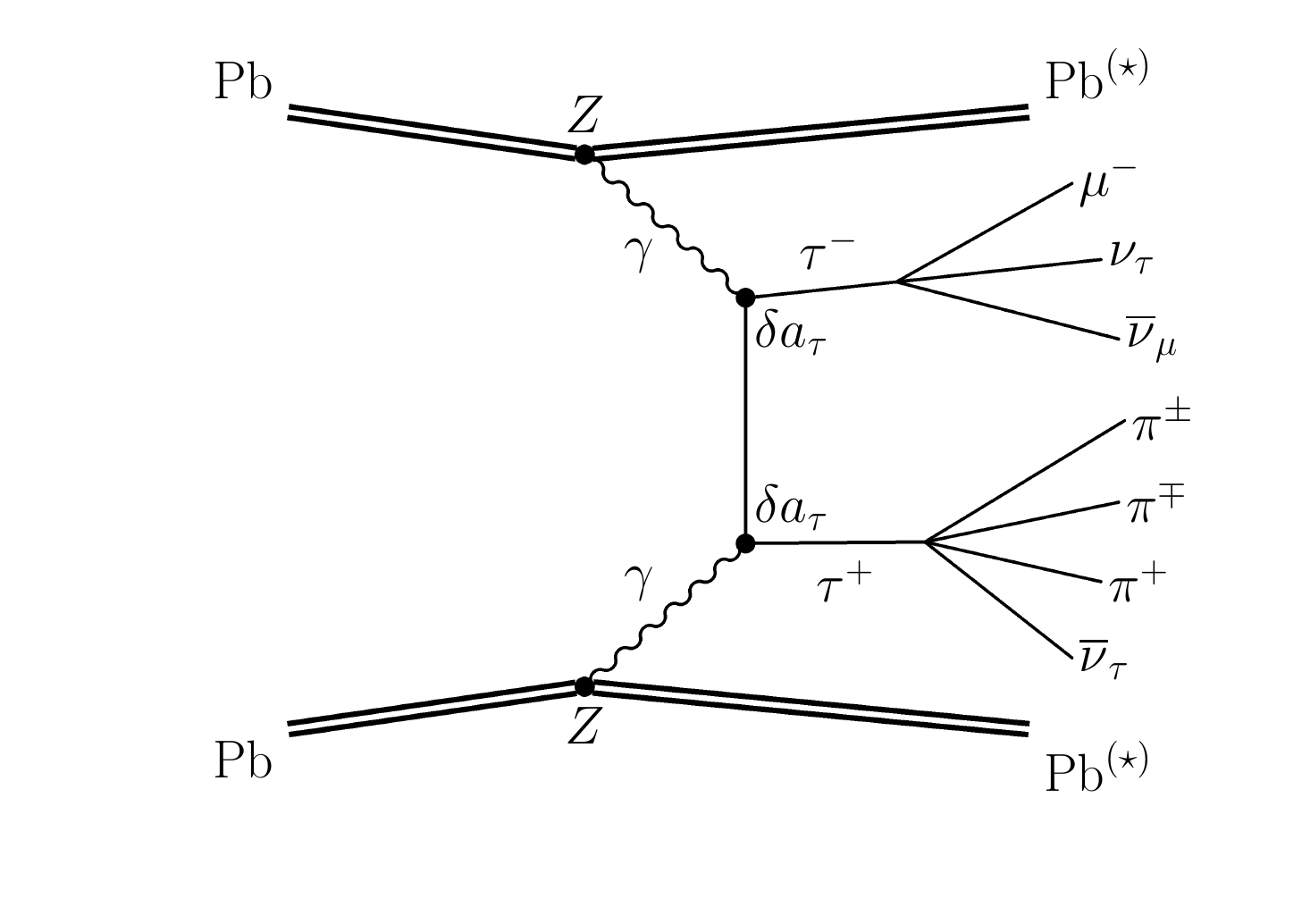}
    \caption{Leading-order QED diagram (and charge conjugate) for the photoproduction of a pair of $\tau$ leptons $\gamma\gamma\to\tau^{+}\tau^{-}$ in PbPb UPC. The presence of $\gamma\tau\tau$ vertices gives sensitivity to the anomalous electromagnetic couplings of the $\tau$ lepton. A possible deviation of the anomalous magnetic moment $\delta a_\tau$ is illustrated in each vertex. The $\tau$ leptons are reconstructed in the final state involving one muon ($\mu$) and three charged particles assumed as pions ($\pi$), while neutrinos ($\nu$) escape undetected. A potential electromagnetic excitation of the outgoing Pb ions is denoted by $(^{\ast})$.~\protect\cite{CMS:2022cji}}
    \label{fig:gtau-schematic}
\end{figure}

Here, we present the first observation of a pair of $\tau$ leptons in PbPb collisions, $\mathrm{Pb}\mathrm{Pb}\,(\gamma\gamma)\to \mathrm{Pb}^{(\ast)}\mathrm{Pb}^{(\ast)}\,\tau^{+}\tau^{-}$ (hereafter referred to as $\gamma\gamma\to\tau^{+}\tau^{-}$). The analysis is based on a data sample collected by the CMS experiment~\cite{CMS:2008xjf} in 2015 at a per nucleon center-of-mass energy $\sqrt{s_{_{\mathrm{NN}}}} = 5.02\,\mathrm{TeV}$, and corresponding to an integrated luminosity of 404$\mu b^{-1}$. As shown schematically in Fig.~\ref{fig:gtau-schematic}, the $\tau$ leptons are reconstructed using the final state of one muon and three charged hadrons (``prongs") assumed to be pions ($\pi$) over a fiducial phase space, defined by the transverse momentum ($p_\mathrm{T}$) and pseudorapidity ($\eta$) of each particle, in order to maximize the signal purity and the detector acceptance and efficiency.

Events are selected in real time by requiring a single muon with no transverse momentum ($p_\mathrm{T}$) threshold requirement, at least one pixel track~\cite{CMS:2014trk}, and a minimum amount of event activity above the noise threshold in the forward hadron (HF) calorimeter. To ensure that events are UPC and further suppress other backgrounds, the energy deposit in the leading tower of HF is required to be below 4$\,$GeV.

In the signal phase space region, one muon and exactly three charged tracks are required. The muon pseudorapidity ($\eta$) is required to be $|\eta|<2.4$, and its $p_\mathrm{T}>3.5\,$GeV for $|\eta|<1.2$ and $p_\mathrm{T}>2.5\,$GeV for $|\eta| \geq 1.2$. The three tracks identified as charged hadrons (pions) and forming the ``$\tau_\mathrm{3prong}$'' lepton candidate~\cite{CMS:2018jrd} are required to be within the tracker acceptance ($|\eta|<2.5$), and to have a common vertex to be within 2.5\,mm of the primary vertex in the $z$ direction. The $p_\mathrm{T}$ must be greater than 0.5 and 0.3$\,$GeV for the leading- and subleading-$p_\mathrm{T}$ pions, respectively. The selected tracks are also required to be labeled as ``high-purity''~\cite{CMS:2014pgm} tracks. The $\tau_\mathrm{3prong}$ candidate is then required to be of opposite charge relative to the selected $\tau_\mu$ candidate, and to have $p_\mathrm{T}^\mathrm{vis}>$2$\,$GeV, where $p_\mathrm{T}^\mathrm{vis}$ is the vector sum $\overrightarrow{p_\mathrm{T}}$ of the three pions. Additionally, the invariant mass of the $\tau_\mathrm{3prong}$ candidate is required to be between 0.2 and 1.5$\,$GeV. The event selection is summarized in Table~\ref{tab:fiducialregion}. 

\begin{table}[t]
\caption[]{The definition of the fiducial phase space region for the $\sigma(\gamma\gamma\to\tau^{+}\tau^{-})$ measurement~\protect\cite{CMS:2022cji}.}
\label{tab:fiducialregion}
\vspace{0.4cm}
\begin{center}
\begin{tabular}{|c|c|}
  \hline
  & \\
  For the $\mu$ & 
$p_\mathrm{T}>3.5\,$GeV for $|\eta|<1.2$    \\ 
                & $p_\mathrm{T}>2.5\,$GeV for $1.2<|\eta|<2.4$\\
& \\
\hline
& \\
  For the pions   & $p_\mathrm{T}^\mathrm{leading}>0.5\,$GeV \& $p_\mathrm{T}>0.3\,$GeV for the (sub-)subleading \\ [0.5ex]
                 & $|\eta|<2.5$  \\ \hline
& \\
  For the $\tau_\mathrm{3prong}$ & $p_\mathrm{T}^\mathrm{vis}>2\,$GeV and $0.2<m_{\tau}^\mathrm{vis}<1.5\,$GeV\\
& \\ \hline
\end{tabular}
\end{center}
\end{table}

The signal is modeled with a dedicated $\gamma\gamma\to\tau^{+}\tau^{-}$ Monte Carlo (MC) sample~\cite{Beresford:2019gww} generated with MADGRAPH5\_aMC@NLO(v2.6.5)~\cite{Alwall:2014hca}, where PYTHIA8~(v2.1.2)~\cite{Sjostrand:2014zea} is used for the hadronization and decay, and GEANT4~\cite{Agostinelli:2002hh} is used to model the detector effects, including resolution, tracking, and trigger efficiencies which are also corrected for by comparing with data. The signal distributions are normalized to match the QED prediction of Ref.~\cite{Beresford:2019gww}. The background is estimated in a data-driven approach using control regions of the phase space with higher number of charged hadron tracks per event or higher energy deposit in HF. Comparing the observed data with the signal simulation and data-driven background we observe good agreement.

\begin{figure}[htbp!]
\centering
\includegraphics[width=0.32\textwidth]{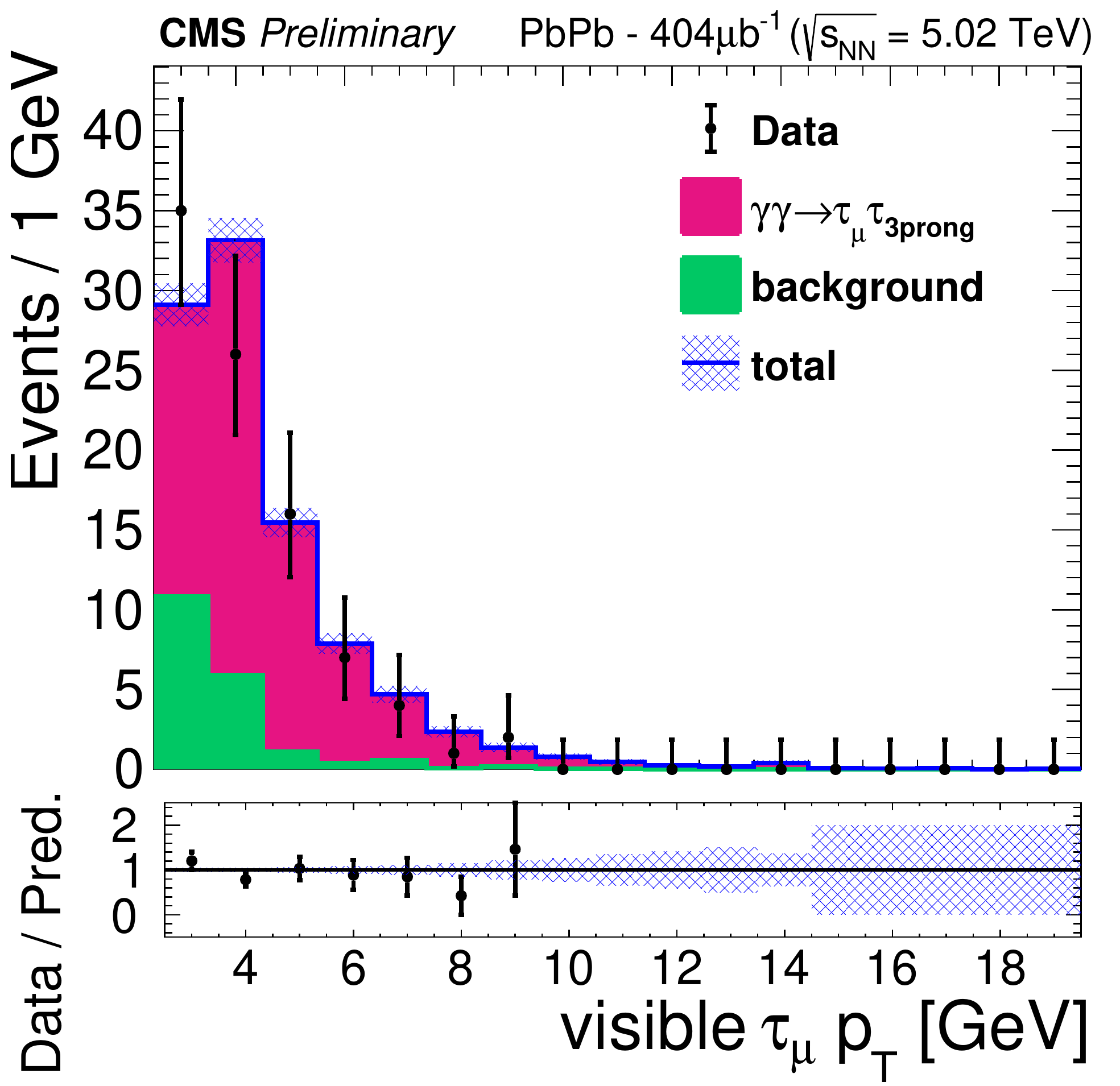}
\includegraphics[keepaspectratio,width=0.32\textwidth]{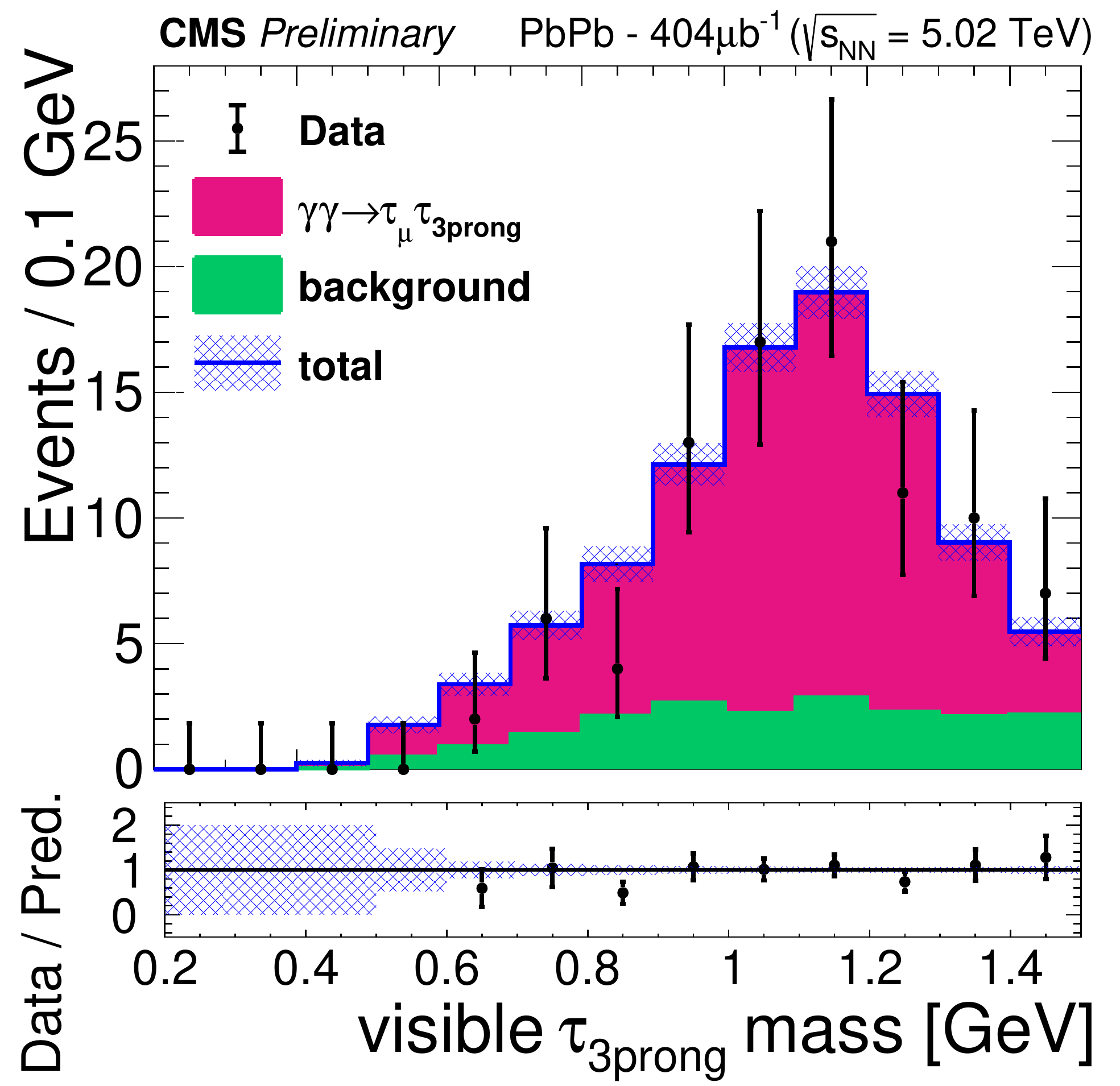}
\includegraphics[keepaspectratio,width=0.32\textwidth]{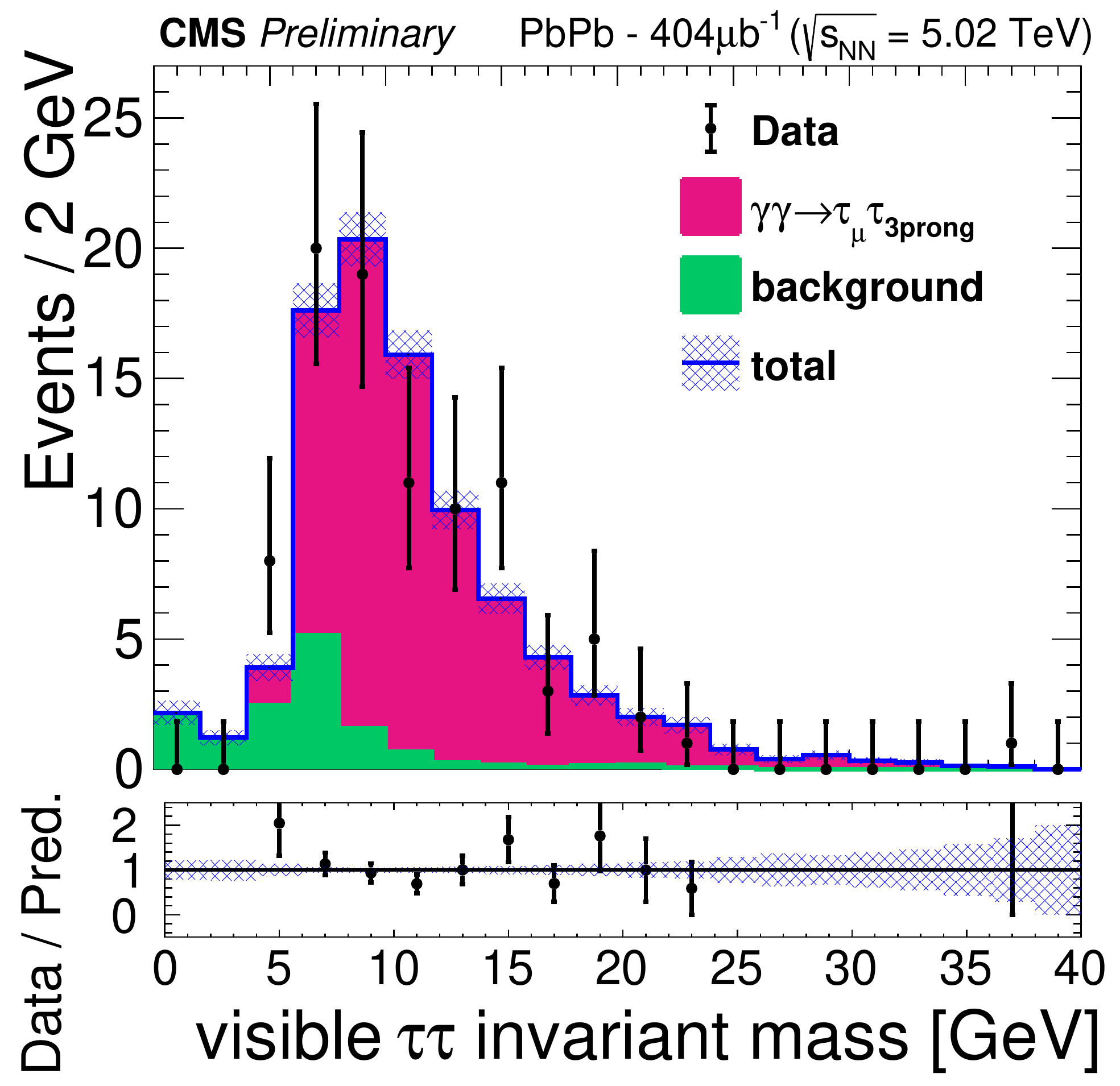}
\caption{Left: Transverse momentum of the muon originating from the $\tau_\mu$ candidate. Middle: Invariant mass of the three pions forming the $\tau_\mathrm{3prong}$ candidate. Right: $\tau\tau$ invariant mass. In all plots, the signal component (magenta histogram) is stacked on top of the background component (green histogram), considering their initial normalization as described in the text. The total is displayed by a blue line and the shaded area shows the statistical uncertainty. The data are represented with black points and the uncertainty is statistical only. The lower panels show the ratios of data to the signal plus background prediction, and the shaded bands represent the statistical uncertainty.~\protect\cite{CMS:2022cji}}
\label{fig:muonKinematicsDataMC}
\end{figure}

The signal is extracted using a binned maximum likelihood fit of signal and background components. The fit is performed on the binned distribution of the difference in azimuthal opening angle between the $\tau_\mu$ and $\tau_\mathrm{3prong}$ candidates, $\Delta\phi(\tau_\mu,\tau_\mathrm{3prong})$. The signal distribution is derived from simulation, while that of the background is obtained from a data-driven method~\cite{CMS:2022cji}. The prefit number of signal events is scaled to match the QED prediction of Ref.~\cite{Beresford:2019gww}. Systematic uncertainties may affect both the normalization and the shape of the $\Delta\phi(\tau_\mu,\tau_\mathrm{3prong})$ distributions. These uncertainties, in addition to the bin-by-bin variations of the signal and background templates, are represented by nuisance parameters in the fit. The negative of the log likelihood is minimized by varying the nuisance parameters according to their uncertainties and by scaling the signal by a multiplicative factor $\mu$.

The systematic uncertainty on the measured cross section coming from a 10\%~\cite{CMS:2018bbk} variation of the HF energy scale is found to be 0.9\%, entirely dominated by the uncertainty in the background shape. As the background shape depends on the high $n_\mathrm{ch}$ parameter as well, an additional uncertainty of 0.2\% on the cross section measurement is evaluated by setting this parameter to 5, 6, 7, and 8 and evaluate the maximum variation with respect to the baseline. The uncertainty in the $\tau$ lepton branching fraction measurements accounts for $0.6\%$~\cite{ParticleDataGroup:2020ssz}. The uncertainty in the muon reconstruction SFs, including the trigger response, identification and tracking efficiency, has an impact of 6.7\%. The uncertainty in the pion tracking SF results in an uncertainty of 3.6\%. The uncertainty in the integrated luminosity amounts to 5.0\%\,\cite{CMS:2021xjt} and affects the normalization of the signal process that is based on simulation. Finally, uncertainties are included from the finite MC sample size that changes the efficiency by 1.1\%, when calculated from the weighted binomial uncertainty, and by 3\%, when allowing for bin-by-bin statistical variations of the MC distribution shape. Summing these uncertainties in quadrature, while taking into account their correlation, a total systematic uncertainty of 9.7\% is found.

\begin{figure}
\centering
\includegraphics[keepaspectratio,width=0.5\textwidth]{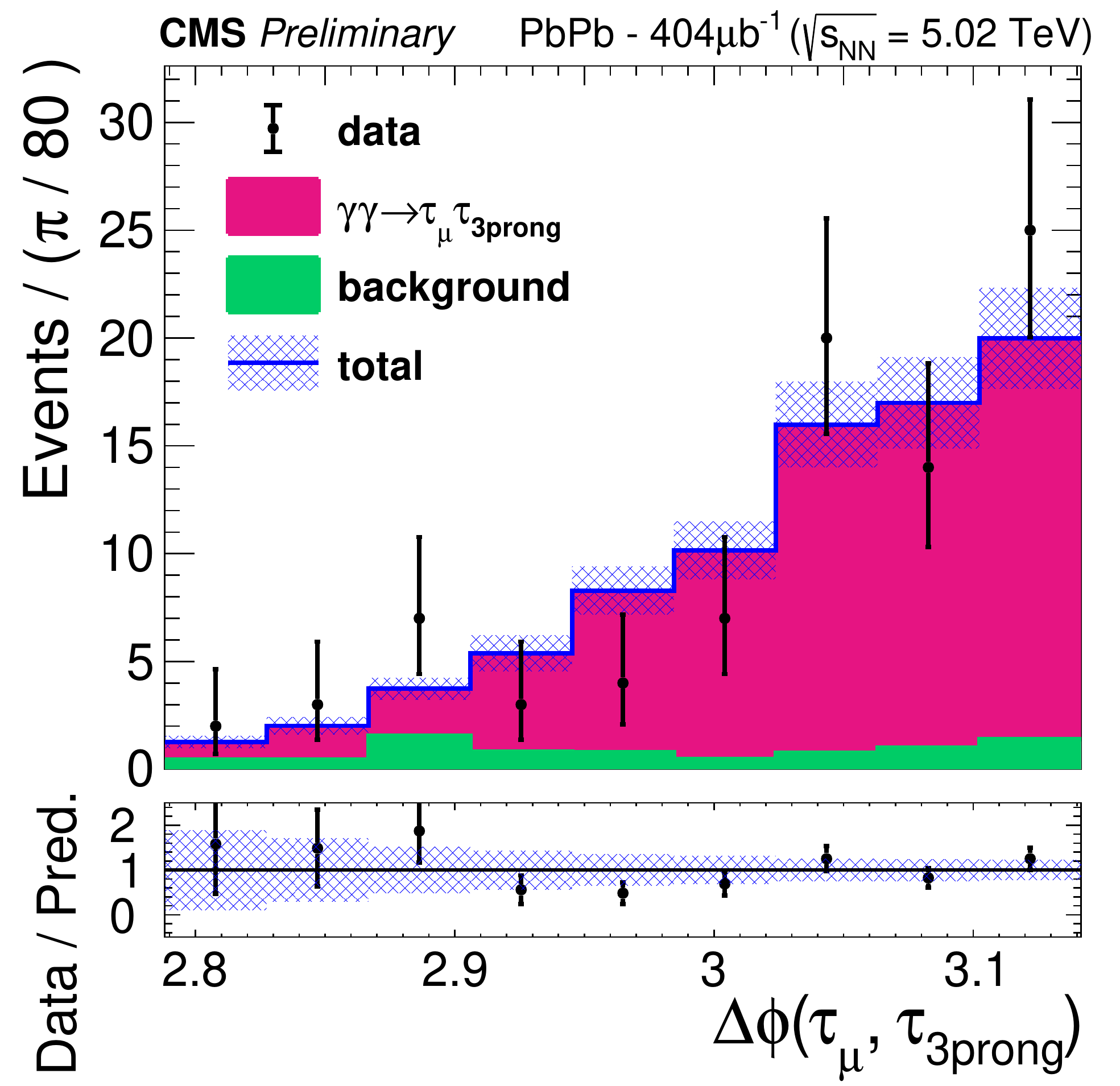}
\caption{Difference in azimuthal opening angle between the $\tau_\mu$ and $\tau_\mathrm{3prong}$ candidates. The data are represented by the points with the vertical bars showing the statistical uncertainties. The signal (background) contribution is given by the magenta (green) histogram, after the application of the fit procedure. The total is displayed by a blue line and the shaded area shows the total uncertainty. The lower panel shows the ratio of data to the signal plus background prediction, and the shaded band represents the total uncertainty.~\protect\cite{CMS:2022cji}}
\label{fig:postfit}
\end{figure}

\begin{figure}[!h!tbp]
  \begin{center}
    \includegraphics[width=0.6\textwidth]{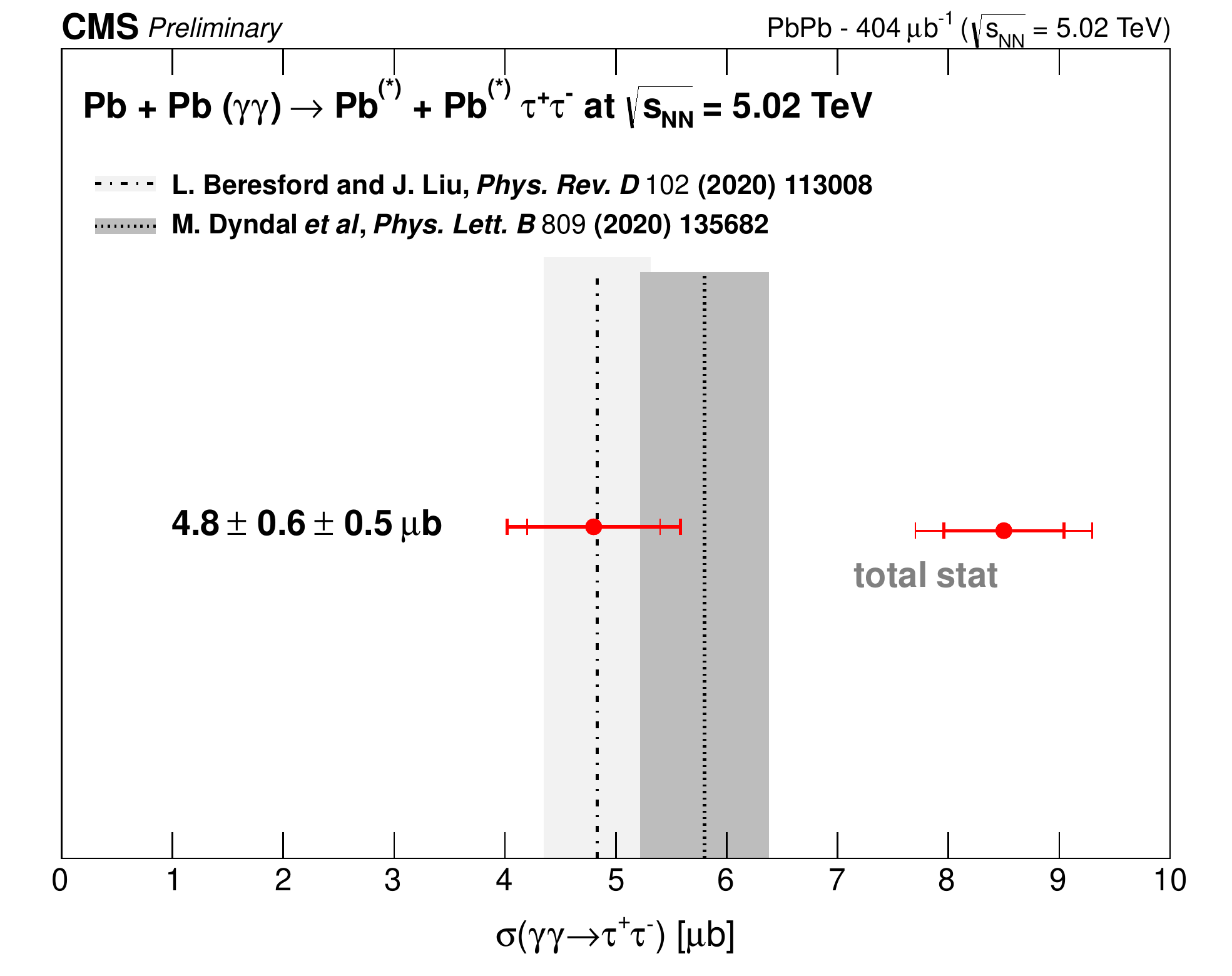}
    \caption{The fiducial cross section, $\sigma(\gamma\gamma\to\tau^{+}\tau^{-})$, measured at $\sqrt{s_{_{\mathrm{NN}}}} = 5.02\,$TeV~\protect\cite{CMS:2022cji}.
    The theoretical predictions are computed with leading-order accuracy in QED and represented by the light~\protect\cite{Beresford:2019gww} and dark~\protect\cite{Dyndal:2020yen} gray bands. 
    }
    \label{fig:sum}
\end{center}
\end{figure}

The best fit value of the signal strength is given by the minimization of the negative log likelihood, and corresponds to $\mu=0.99^{+0.16}_{-0.14}$ with $N^{\tau\tau}_\mathrm{sig}=77\pm 12$ signal events in the integral of the postfit signal component. The fit result is shown in Fig.~\ref{fig:postfit}, where the signal template is represented by the magenta histogram, the background by the green histogram, and the data by the black points. The contributions are stacked with their total uncertainty represented by the blue hatched area. The fiducial cross section is found to be $\sigma(\gamma\gamma\to\tau^{+}\tau^{-}) = 4.8\pm 0.6\,(\mathrm{stat})\pm 0.5\,(\mathrm{syst})~\mu\mathrm{b}$. The result, summarized in Fig.~\ref{fig:sum}, is compared to leading-order QED predictions~\cite{Beresford:2019gww,Dyndal:2020yen}. The significance we obtain from the fit is greater than 5 standard deviations.

\begin{figure}[!h!tbp]
  \begin{center}
    \includegraphics[width=0.78\textwidth]{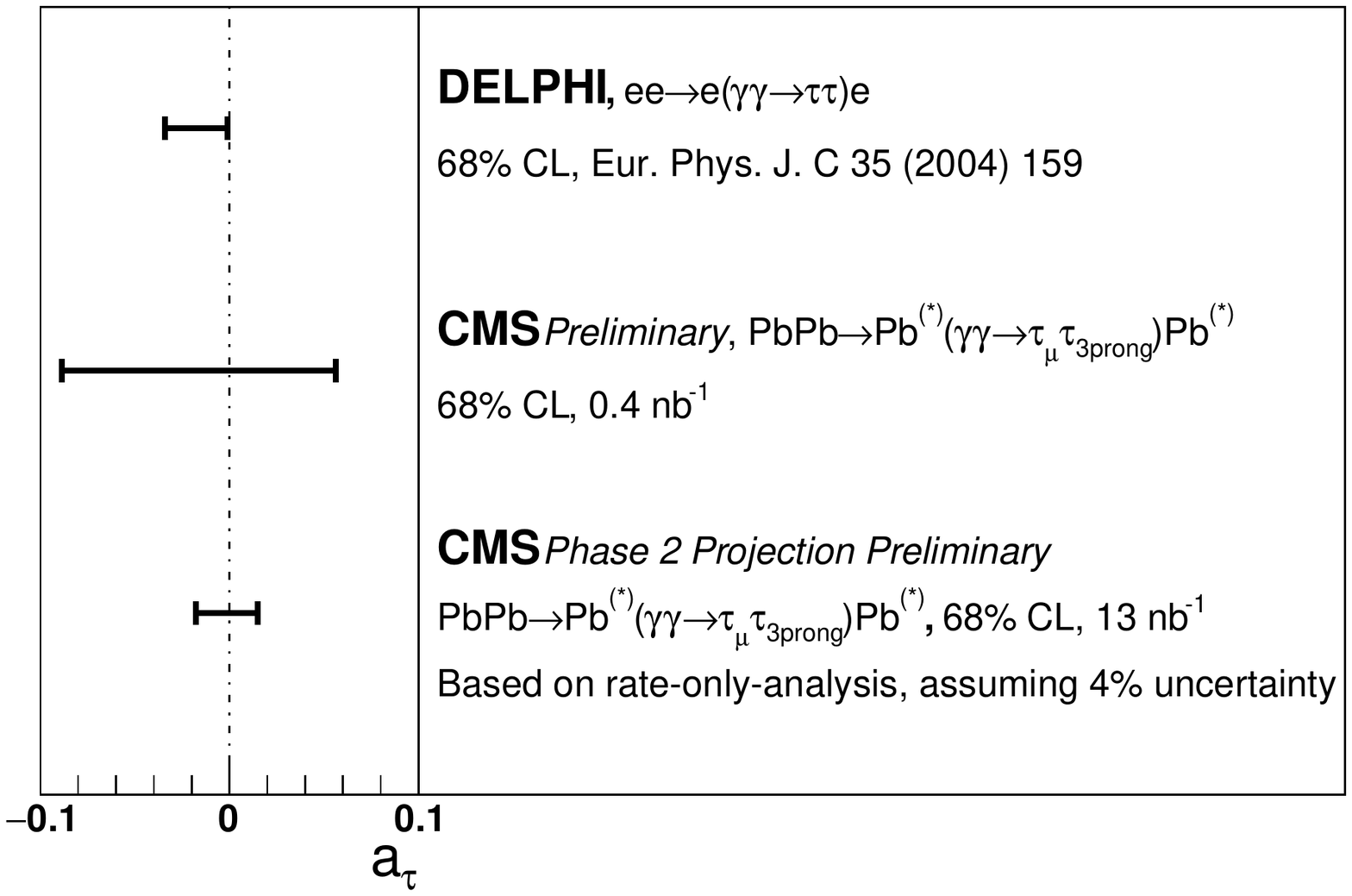}
    \caption{Comparison of the constraints on $a_\tau$ at 68\% confidence level from  this analysis~\protect\cite{CMS:2022cji} and the DELPHI experiment at LEP~\protect\cite{DELPHI:2003nah}. The projection to the integrated PbPb luminosity expected from Runs 3 and 4 of the LHC nuclear program is included as well. For the latter, we foresee a $<$4 ($<$2)\,\% systematic (statistical) uncertainty with the improvements originating from lepton and tracking reconstruction, and the knowledge of the luminosity.}
    \label{fig:atau}
\end{center}
\end{figure}

 We use variations of the total $\sigma(\gamma\gamma\to\tau^{+}\tau^{-})$ to place model-dependent~\cite{Beresford:2019gww}, first limits on $a_\tau$ at the LHC.
These limits are $(-8.8<a_\tau<5.6)\times 10^{-2}$ at 68\% confidence level and can be compared with the current best measurement from the DELPHI Collaboration~\cite{DELPHI:2003nah} of $a_\tau=(-1.8\pm 1.7)\times 10^{-2}$. A comparison of this measurement to the current world's best and a projection to the expected integrated luminosity for Runs 3 and 4 of the LHC is shown in Fig.~\ref{fig:atau}

In summary, the first observation of $\tau$ lepton pair production in ultraperipheral nucleus-nucleus collisions is reported. Events with the final state of one muon and three charged particles identified as pions are reconstructed from a lead-lead data sample collected by the CMS experiment at $\sqrt{s_{_{\mathrm{NN}}}} = 5.02\,$TeV in 2015, and corresponding to an integrated luminosity of 404$\,\mu\mathrm{b}^{-1}$. The statistical significance of the signal relative to the background-only expectation is above five standard deviations. The cross section for the $\gamma\gamma\to\tau^{+}\tau^{-}$ process, within a fiducial phase space region, is $4.8\pm 0.6\,(\mathrm{stat})\pm 0.5\,(\mathrm{syst})~\mu\mathrm{b}$, in agreement with predictions from quantum electrodynamics at leading-order accuracy. This measurement  introduces a novel experimental strategy using heavy ion collisions already recorded by the LHC, which is expected, with the incorporation of additional data, to surpass the precision on the $\tau$ magnetic moment attained previously at lepton-lepton colliders. Using the measured cross section and its corresponding uncertainties, we estimate a first limit of $(-8.8<a_\tau<5.6)\times 10^{-2}$ with 68\% confidence level at the LHC.


\section*{References}

\begin{thebibliography}{99}


\bibitem{Baltz:2008}B. A. Baltz and G. Baur and D. d'Enterria and L. Frankfurt and F. Gelis, {\em Phys. Rept.} {\bf 458}, 1 (2008)

\bibitem{Beresford:2019gww}L. Beresford and J. Liu, \Journal{\PRD}{102}{113008}{2020}

\bibitem{Dyndal:2020yen}M. Dyndal, M. Klusek-Gawenda, M. Schott, and A. Szczurek, \Journal{\PLB}{809}{135682}{2020}

\bibitem{Crivellin:2021spu}A. Crivellin and M. Hoferichter and J. M. Roney, {\em arXiv:2111.10378}

\bibitem{CMS:2008xjf}CMS Collaboration, {\em JINST} {\bf 3}, S08004 (2008)

\bibitem{CMS:2014trk}CMS Collaboration, {\em JINST} {\bf 9}, P10009 (2014)

\bibitem{CMS:2022cji}CMS Collaboration, \href{http://cds.cern.ch/record/2803742}{CMS-PAS-HIN-21-009} (2022)

\bibitem{CMS:2018jrd}CMS Collaboration, {\em JINST} {\bf 13}, P10005 (2018)

\bibitem{CMS:2014pgm}CMS Collaboration, {\em JINST} {\bf 9}, P10009 (2014)

\bibitem{Alwall:2014hca}J. Alwall, R. Frederix, S. Frixione, V. Hirschi, F. Maltoni, O. Mattelaer, H.-S. Shao, T. Stelzer, P. Torrielli, and M. Zaro, {\em JHEP} {\bf 7}, 79 (2014)

\bibitem{Sjostrand:2014zea}T. Sj{\"o}strand, S. Ask, J. R. Christiansen, R. Corke, N. Desai, P. Ilten, S. Mrenna, S. Prestel, C. O. Rasmussen, and P. Skands, {\em Comput. Phys. Commun.} {\bf 191}, 159 (2015)

\bibitem{Agostinelli:2002hh}GEANT4 Collaboration, {\em Nucl. Instrum. Meth.} A {\bf 506}, 250 (2003)

\bibitem{CMS:2018bbk}CMS Collaboration, {\em Eur. Phys. J.} C {\bf 79}, 277 (2019)

\bibitem{ParticleDataGroup:2020ssz}Particle Data Group, {\em PTEP} {\bf 2020}, 083C01 (2020)

\bibitem{CMS:2021xjt}CMS Collaboration, {\em Eur. Phys. J.} C {\bf 81}, 800 (2021)

\bibitem{DELPHI:2003nah}DELPHI Collaboration, {\em Eur. Phys. J.} C {\bf 35}, 159 (2004)

\end{thebibliography}
\end{document}